%% file: maxentropy.tex
\documentclass[conference]{IEEEtran}
\usepackage{amsmath}
\usepackage{amsthm}
\usepackage{amssymb}
\usepackage{bm}
\usepackage{xspace}
\usepackage{xcolor}
\usepackage{graphicx}
\usepackage{psfrag}
\usepackage{url}
\usepackage{framed}
\usepackage{cite}

\usepackage{flushend}

\DeclareMathOperator{\entop}{H}
\DeclareMathOperator{\expop}{E}
\DeclareMathOperator{\probop}{P}
\DeclareMathOperator{\gf}{G}
\DeclareMathOperator{\repart}{Re}
\DeclareMathOperator{\catop}{cat}

\newcommand{\stra}{\mathtt{a}}
\newcommand{\strb}{\mathtt{b}}
\newcommand{\strzero}{\mathtt{0}}
\newcommand{\strone}{\mathtt{1}}

\renewcommand{\figurename}{Example}

\newtheorem{theorem}{Theorem}
\newtheorem{lemma}{Lemma}
\theoremstyle{definition}
\newtheorem{definition}{Definition}

\title{On the Capacity of Constrained Systems}

\IEEEoverridecommandlockouts

\author{\IEEEauthorblockN{G. B\"ocherer,
R. Mathar}
\IEEEauthorblockA{\IEEEauthorrefmark{1}Institute for Theoretical Information
Technology\\
RWTH Aachen University\\
52056 Aachen, Germany\\ Email: \{boecherer,mathar\}@ti.rwth-aachen.de}
\and
\IEEEauthorblockN{V.C. da Rocha Jr.,
C. Pimentel}
\IEEEauthorblockA{
Communications Research Group - CODEC\\
Department of Electronics and Systems, P.O. Box 7800\\
Federal University of Pernambuco\\
50711-970 Recife PE, Brazil\\
E-mail: \{vcr,cecilio\}@ufpe.br
}
\thanks{This work has been supported by the UMIC Research Centre, RWTH
Aachen University. This work has also been supported by the Brazilian National Council for Scientific and Technological Development (CNPq).}
}

\newcounter{textequation}
\newcounter{exampleequation}
\begin{document}

\maketitle
\input{abstract}

\input{introduction}

\input{constrained}

\input{combinatorial}

\input{relation}

\input{maxentropic}

\input{conclusions}

\bibliographystyle{IEEEtran}
\bibliography{IEEEabrv,confs-jrnls,Literatur}

\end{document}

%% file: abstract.tex
\begin{abstract}
In the first chapter of Shannon's \emph{A Mathematical Theory of Communication}, it is shown that
the maximum entropy rate of an input process of a constrained system is limited by
the combinatorial capacity of the system. Shannon considers systems where the constraints define 
regular languages and uses results from matrix theory in his derivations. In this work, the regularity constraint is dropped. Using generating functions, it is shown that the maximum entropy rate of an input process is upper-bounded by the combinatorial capacity in general. The presented results also allow for a new approach to systems with regular constraints. As an example, the results are applied to binary sequences that fulfill the $(j,k)$ run-length constraint and by using the proposed framework, a simple formula for the combinatorial capacity is given and a maxentropic input process is defined.
\end{abstract}

%% file: introduction.tex
\section{Introduction}

This work is motivated by the recent interest in the information-theoretic limits of systems with constraints that do not form a regular language. One example is the consideration of context-free languages with an application to genetic sequence modelling \cite{Milenkovic2007}, another example is the investigation of asynchronous channels \cite{Yeung2009}.

A constrained system allows the noiseless transmission of input sequences of weighted symbols that fulfill certain constraints on the symbol constellations. A natural question is how to efficiently encode a random source such that it becomes a valid input for a constrained system \cite{Marcus2001}. Furthermore, it is of interest to determine the ultimate performance of such an encoder, which is closely related to the entropy rate of random processes that generate strings that fulfill the constraints. 
The maximum entropy rate of all such processes is equal to the combinatorial capacity of the considered system in the case that the constraints form a regular language. This was originally shown in \cite{Shannon1948}. In \cite{Khandekar2000}, the authors show this property for a slightly generalized setup, since they allow non-integer valued symbol weights, as long as the set of weights is not too dense. We will define what ``not too dense'' means in Section~\ref{sec:combinatorial}. Recently, the authors of \cite{Yeung2009} showed that combinatorial capacity and maximum entropy rate are equal for a specific class of constrained systems, which they call the asynchronous channel. It is worthy to note that for asynchronous channels, the ``not too dense'' assumption is not necessarily fulfilled.

In this work, we consider constrained systems with not necessarily regular constraints; we allow symbol weights taking arbitrary positive real values, and motivated by \cite{Yeung2009}, we allow the set of symbol weights to be too dense. For this general class of constrained systems, we show how to represent such systems by generating functions. We give a new definition of combinatorial capacity that coincides with the original definition \cite{Shannon1948} when the weight set is not too dense. By invoking known results, we show that the combinatorial capacity of such systems is equal to the abscissa of convergence of the corresponding generating function. Finally, we define input processes of constrained systems and show that the maximum entropy rate of input processes is upper-bounded by the abscissa of convergence of the generating function. This is our main result: independent from if the ``not too dense'' property is fulfilled or not, and for whatever constraints, the combinatorial capacity is equal to the abscissa of convergence of the generating function, and the entropy rate is upper-bounded by the abscissa of convergence of the generating function. By a detailed discussion of the $(j,k)$ run-length constraint, we illustrate our ideas and we show that our framework, besides being more general, also allows for a new approach to investigate regular systems. Namely, we derive a simple formula for the combinatorial capacity of the $(j,k)$ constraint and we define an input process whose entropy rate is equal to the combinatorial capacity.

The remainder of this paper is organized as follows. In Section~\ref{sec:constrainedSystems}, we define constrained systems and generating functions. In Section~\ref{sec:combinatorial}, we show how to calculate the combinatorial capacity. We then show in Section~\ref{sec:relation} how the combinatorial capacity relates to entropy rates and finally, in Section~\ref{sec:inputProcess}, we show that the entropy rate of input processes is upper-bounded by the combinatorial capacity.

%% file: constrained.tex
\section{Constrained Systems}
\label{sec:constrainedSystems}
%
%
\begin{figure*}
\setcounter{textequation}{\value{equation}}
\renewcommand{\theequation}{E.\arabic{equation}}
\setcounter{equation}{\value{exampleequation}}
\begin{minipage}{0.99\textwidth}
\begin{framed}
A system $S_{(j,k)}=(\mathcal{A},w)$ accepting binary strings that fulfill the $(j,k)$ constraint can be defined as follows using regular expressions \cite{Sipser2006}:
\begin{align}
\mathcal{A}&=(\strone\cup\dotsb\cup \underbrace{\strone\dotsb \strone}_\text{$j$ times})[(\strzero\cup\dotsb\cup \underbrace{\strzero\dotsb \strzero}_\text{$k$ times})(\strone\cup\dotsb\cup \underbrace{\strone\dotsb \strone}_\text{$j$ times})]^\star(\varepsilon\cup \strzero\cup\dotsb\cup \underbrace{\strzero\dotsb \strzero}_\text{$k$ times})
\nonumber\\
&\qquad\cup(\strzero\cup\dotsb\cup \underbrace{\strzero\dotsb \strzero}_\text{$k$ times})[(\strone\cup\dotsb\cup \underbrace{\strone\dotsb \strone}_\text{$j$ times})(\strzero\cup\dotsb\cup \underbrace{\strzero\dotsb \strzero}_\text{$k$ times})]^\star(\varepsilon\cup \strone\cup\dotsb\cup \underbrace{\strone\dotsb \strone}_\text{$j$ times})\label{eq:example:regular}\\
w(\strzero)&=w(\strone)=1.\label{eq:example:weight}
\end{align}
The symbol $^\star$ denotes the Kleene star, $\cup$ denotes the or-operation, $\varepsilon$ denotes the empty-string with $w(\varepsilon)=0$, $\strone\strone$ denotes ``$\strone$ concatenated with $\strone$''. Note that concatenation is non-commutative: $\strone(\strone\cup\strzero)=\strone\strone\cup\strone\strzero\neq\strone\strone\cup\strzero\strone$.
\end{framed}
\vspace{-0.5cm}
\caption{Definition of $S_{(j,k)}$.}
\label{fig:constrainedSystem}
\end{minipage}
\\[0.2cm]
\begin{minipage}{0.99\textwidth}
\begin{framed}
From \eqref{eq:example:regular} and \eqref{eq:example:weight} in Example~\ref{fig:constrainedSystem}, the generating function of $S_{(j,k)}$ can directly be derived as 
\begin{align}
\gf_{(j,k)}(s)&=(e^{-s}+\dotsb+e^{-js})\sum\limits_{n=0}^\infty[(e^{-s}+\dotsb+e^{-ks})(e^{-s}+\dotsb+e^{-js})]^n(1+e^{-s}+\dotsb+e^{-ks})\nonumber\\
&+(e^{-s}+\dotsb+e^{-ks})\sum\limits_{n=0}^\infty[(e^{-s}+\dotsb+e^{-js})(e^{-s}+\dotsb+e^{-ks})]^n(1+e^{-s}+\dotsb+e^{-js}).
\label{eq:example:generatingFunction}
\end{align}
A discussion of how to derive generating functions from regular expressions can be found in \cite{Bocherer2007a}.
\end{framed}
\vspace{-0.4cm}
\caption{Generating function of $S_{(j,k)}$.}
\vspace{-0.4cm}
\label{fig:generatingFunction}
\end{minipage}
\setcounter{exampleequation}{\value{equation}}
\setcounter{equation}{\value{textequation}}
\end{figure*}
%
%
In this section, we define the class of constrained systems that we will investigate in this work and we show how to represent them by generating functions.
\begin{definition}
  A \emph{constrained system} $A=(\mathcal{A},w)$ consists of a countable set $\mathcal{A}$ of strings
  accepted by the system and an associated weight function $w\colon
  \mathcal{A}\rightarrow\mathbb{R}_{>0}$ ($\mathbb{R}_{>0}$ denotes the positive real numbers) with the following property: if $\stra,\strb\in\mathcal{A}$ and $\stra\strb\in\mathcal{A}$ then $w(\stra\strb)=w(\stra)+w(\strb)$. 
\end{definition}
The weight of a symbol can have different practical meaning. In the context of magnetic recording systems, ``weight'' will probably refer to ``tape-length''; other meanings like ``time'' or ``energy'' are possible, depending on the modelled system. For an illustration of our definition, we define in Example~\ref{fig:constrainedSystem} the system $S_{(j,k)}$, which accepts binary sequences that consist of at most $j$ consecutive $\strone$s and at most $k$ consecutive $\strzero$s. This constrained is called the $(j,k)$ run-length constraint. 
We return to $S_{(j,k)}$ at several points in our paper. 
\subsection{Generating Functions}
To analyze the asymptotic behavior of constrained systems, we first represent the set $\mathcal{A}$ of allowed strings together with the weight function $w$ by a generating function. We then interpret the generating function as a function on the complex plane and investigate its convergence behavior. This approach, mostly referred to as
\emph{analytic combinatorics}, is discussed in detail in \cite{Flajolet2008}. We
consider a more general case since we do not restrict the range of the weight function $w$ to the natural numbers, but allow for the set of positive real numbers $\mathbb{R}_{>0}$. Therefore, we use general Dirichlet series \cite{Hardy1915} instead of power series as generating functions.
\begin{definition}\label{def:generatingFunction}
  Let $A=(\mathcal{A},w)$ represent a constrained system. We define the \emph{generating
    function} of $A$ by
  \begin{align}
    \gf_A(s)&=\sum\limits_{a\in \mathcal{A}}e^{-w(a)s},\qquad s\in\mathbb{C}
  \end{align}
where $\mathbb{C}$ denotes the set of complex numbers.
\end{definition}
Let  $\Omega$ denote the set of distinct string weights of elements in $\mathcal{A}$. We
order and index the set $\Omega$ such that
$\Omega=\lbrace \nu_k\rbrace_{k=1}^{\infty}$ with
\mbox{$\nu_1<\nu_2<\dotsb$}. For every $\nu_k\in\Omega$,
$N(\nu_k)$ denotes the number of distinct strings of weight $
\nu_k$ in $\mathcal{A}$. We can now write the generating function as
\begin{align}
  \gf_A(s)=\sum\limits_{k=1}^\infty N(\nu_k)e^{-\nu_k s}.
\end{align}
Since the coefficients $N(\nu_k)$ result from an enumeration, they are all
non-negative. In Example~\ref{fig:generatingFunction}, we show how to represent $S_{(j,k)}$ by a generating function.

%% file: combinatorial.tex
\section{Combinatorial Capacity}
\label{sec:combinatorial}
In previous works that consider non-integer valued weights \cite{Khandekar2000,Bocherer2007}, the authors restrict themselves to constrained systems
 where the ordered set of string weights $\Omega=\lbrace \nu_k\rbrace_{k=1}^{\infty}$ is \emph{not too
dense}, that is, there exists some constant $L\geq 0$ and some constant $K\geq 0$ such that for any integer $n\geq 0$
\begin{align}
  \max_{\nu_k<n} k\leq Ln^K\label{eq:notTooDense}.
\end{align}
Under the ``not too dense'' assumption, it is meaningful to follow Shannon's original definition and to identify the combinatorial capacity with $\mathsf{C}_0$ given by
\begin{align}
\mathsf{C}_0=\underset{k\rightarrow\infty}{\limsup}\dfrac{\ln N(\nu_k)}{\nu_k}\label{eq:conventionalCombinatorial}
\end{align}
as was done for instance in \cite{Khandekar2000} and \cite{Bocherer2007}. Here and hereafter, $\ln$ denotes the natural logarithm. Throughout the paper, for a sequence $\{s_l\}_{l=1}^\infty$,
 $s=\limsup_{l\rightarrow\infty}s_l$ 
is equivalent to the following: for any $\epsilon>0$, it holds that
\begin{align}
 s_l&\leq s+\epsilon\quad\text{almost everywhere (a.e.)}\\
&\text{and}\nonumber\\
 s_l&\geq s-\epsilon\quad\text{infinitely often (i.o.)}
\end{align}
with respect to $l\in\mathbb{N}$ ($\mathbb{N}$ denotes the set of natural numbers starting with one).

If $\Omega$ is too dense, the number of possible string weights in the interval $[n,n+1]$
increases faster than polynomial with $n$, in which case identifying combinatorial capacity
with $\mathsf{C}_0$ may become inappropriate. See \cite{Bocherer2007} for an example and
see \cite{Khandekar2000} for a detailed discussion.

We now give a definition of combinatorial capacity that is meaningful also when the ordered set of string weights $\Omega$ is too dense.
\begin{definition}\label{def:combinatorialCapacity}
We define the \emph{combinatorial capacity} $\mathsf{C}$ as
\begin{align}
\mathsf{C}=\underset{k\rightarrow\infty}{\limsup}\dfrac{\ln \bigl[\sum\limits_{l=1}^k N(\nu_l)\bigr]}{\nu_k}.
\label{eq:combinatorialCapacity}
\end{align}
\end{definition}
The motivation for our generalized definition is twofold. First, our results on entropy rates for constrained systems, which we will present in the remaining sections, do not depend on the ``not too dense'' property. Second, recent work \cite{Yeung2009} has shown that there are constrained systems of practical interest that may not necessarily have the ``not too dense'' property. The following theorem shows that Definition~\ref{def:combinatorialCapacity} of combinatorial capacity is consistent with the conventional definition \eqref{eq:conventionalCombinatorial}, namely, if $\Omega$ is not too dense, then our definition of combinatorial capacity coincides with \eqref{eq:conventionalCombinatorial}, i.e., $\mathsf{C}=\mathsf{C}_0$. 
\begin{theorem}\label{theo:capacityConvergence}
Let $A=(\mathcal{A},w)$ be a constrained system with the set of distinct string weights $\Omega$ and generating function $\gf_A(s)$. Denote the abscissa of convergence of $\gf_A$ by $Q$. The following holds:
\begin{enumerate}
\item The combinatorial capacity $\mathsf{C}$ is equal to $Q$, i.e., 
\begin{align}
\underset{k\rightarrow\infty}{\limsup}\dfrac{\ln \bigl[\sum\limits_{l=1}^k N(\nu_l)\bigr]}{\nu_k}=Q.
\end{align}
\item If the set of distinct weights $\Omega$ is not too dense, then $\mathsf{C}_0=\mathsf{C}$, i.e.,
\begin{align}
\underset{k\rightarrow\infty}{\limsup}\dfrac{\ln N(\nu_k)}{\nu_k}=\underset{k\rightarrow\infty}{\limsup}\dfrac{\ln \bigl[\sum\limits_{l=1}^k N(\nu_l)\bigr]}{\nu_k}.
\end{align}
\end{enumerate}
\end{theorem}
\begin{IEEEproof}
We start by proving statement (i). All coefficients $N(\nu_k)$ are non-negative since they result from an enumeration. Therefore, for all $k\in\mathbb{N}$, $N(\nu_k)=|N(\nu_k)|$. With this observation, (i) follows directly from \cite[Theorem 7]{Hardy1915}. To proof statement (ii), we assume that $\Omega$ is not too dense. In this case $\mathsf{C}_0=Q$, which was shown in \cite[Lemma 1]{Bocherer2007}. Statement (ii) now follows from statement (i). Alternatively, (ii) can be shown by combinatorial arguments, see \cite[Appendix B.3]{Bocherer2007a}
\end{IEEEproof}
Returning to our example, the set of symbol weights of $S_{(j,k)}$ is not too dense, because the underlying alphabet $\{\strzero,\strone\}$ of $\mathcal{A}$ is finite. See \cite[Appendix A]{Khandekar2000} for a more detailed discussion of this argument. Using Theorem~\ref{theo:capacityConvergence}, we derive in Example~\ref{fig:combinatorialCapacity} a simple formula for the combinatorial capacity of $S_{(j,k)}$.
%
%
\begin{figure}
\setcounter{textequation}{\value{equation}}
\renewcommand{\theequation}{E.\arabic{equation}}
\setcounter{equation}{\value{exampleequation}}
\begin{framed}
By Theorem~\ref{theo:capacityConvergence}, the combinatorial capacity of $S_{(j,k)}$ is given by the abscissa of convergence of its generating function $\gf_{(j,k)}$. From \eqref{eq:example:generatingFunction} in Example~\ref{fig:constrainedSystem}, we see that the abscissa of convergence of $\gf_{(j,k)}$  is given by the largest positive real solution of
\begin{align}
(e^{-s}+\dotsb+e^{-js})(e^{-s}+\dotsb+e^{-ks})=1.\label{eq:exampleCapacity}
\end{align}
This formula coincides with the formula given in \cite[Theorem 2]{Mittelholzer2009} and it can also be derived by applying the techniques introduced in \cite{Pimentel2003}.
\end{framed}
\vspace{-0.4cm}
\caption{Combinatorial capacity of $S_{(j,k)}$.}
\vspace{-0.4cm}
\label{fig:combinatorialCapacity}
\setcounter{exampleequation}{\value{equation}}
\setcounter{equation}{\value{textequation}}
\end{figure}
%
%

%% file: relation.tex
\section{The Relation Between Combinatorial Capacity and Entropy Rates}
\label{sec:relation}
After having defined the combinatorial capacity of constrained systems in the last section, we now want to consider random processes that generate strings that fulfill the constraints of the considered system. We then want to know how the maximum entropy rate of such a process relates to the combinatorial capacity. Ultimately, we have a process in mind that generates at each time instant a substring, which is then appended to the string that has been generated so far, such that at each time instant, the generated string fulfills the constraints of the system. In magnetic recording, such a process would generate a substring, write it to the tape, generate another substring, write it to the tape, and so forth, without ever rewinding the tape. The difficulty of analyzing such a process is the following: fix two time instants $l$ and $l'$, $l<l'$. The probability that the process writes a specific string to the tape until time instant $l'$ depends in general on the probabilities of the strings that it can write until time instant $l$. This dependency can become arbitrarily complicated depending on the constraints of the system. Because of these interdependencies, it is difficult to bound the entropy rate of such a process. We solve these interdependencies by decoupling the time instants $l$ and $l'$: each time the recording system wants to write to the tape, it first rewinds the tape completely and then overwrites everything that has been written before. We call such a system an input source (in contrast to an input process) of a constrained system. In this section, we show through a series of results that the entropy rate of input sources is upper-bounded by the combinatorial capacity and we postpone input processes until Section~\ref{sec:inputProcess}.
\subsection{Input Sources for Constrained Systems}
\begin{definition}\label{def:inputSource}
Let $A=(\mathcal{A},w)$ denote a constrained system. Denote by $X=\{X_l\}_{l=1}^\infty$ a sequence of random variables and denote by $\mathcal{X}_l$ the support of $X_l$. We say that $X$ is an \emph{input source} of $A$ if and only if
\begin{enumerate}
\item $\bigcup_{l=1}^\infty\mathcal{X}_l \subseteq \mathcal{A}$ and $\mathcal{X}_l\cap\mathcal{X}_k=\emptyset$, if $l\neq k$.\label{enum:input1}
\item For each $l\in\mathbb{N}$, $\mathcal{X}_l\neq\emptyset$.\label{enum:input2}
\end{enumerate}
\end{definition}
%
%
\begin{figure}
\setcounter{textequation}{\value{equation}}
\renewcommand{\theequation}{E.\arabic{equation}}
\setcounter{equation}{\value{exampleequation}}
\begin{framed}
The sequence of random variables $\{X_l\}_{l=1}^\infty$ with support of $X_l$ given by
\begin{align}
\mathcal{X}_l=[(\strzero\cup\dotsb\cup \underbrace{\strzero\dotsb \strzero}_\text{$k$ times})(\strone\cup\dotsb\cup \underbrace{\strone\dotsb \strone}_\text{$j$ times})]^l
\end{align}
is an input source of $A$: first, $\bigcup_{l=1}^\infty\mathcal{X}_l\subseteq\mathcal{A}$ and $\mathcal{X}_l\cap\mathcal{X}_k=\emptyset$ whenever $l\neq k$ and second, $\mathcal{X}\neq\emptyset$ for each $l\in\mathbb{N}$, which shows that both condition (i) and condition (ii) of Definition~\ref{def:inputSource} are fulfilled.
\end{framed}
\vspace{-0.4cm}
\caption{An input source for $S_{(j,k)}$.}
\vspace{-0.4cm}
\label{fig:inputSource}
\setcounter{exampleequation}{\value{equation}}
\setcounter{equation}{\value{textequation}}
\end{figure}
%
%
We define in Example~\ref{fig:inputSource} an input source for $S_{(j,k)}$. Note that the given example is not the only possible input source of $S_{(j,k)}$.

We denote the \emph{probability mass function} (PMF) of $X_l$ by
\begin{align}
p_{X_l}(x)=\probop[X_l=x],\quad x\in\mathcal{X}_l.
\end{align}
\begin{definition}
We define the \emph{entropy rate}
$\bar{\entop}$ of an input source $X$ by
\begin{align}
\bar{\entop}(X)=\limsup_{l\rightarrow\infty}\frac{\entop(X_l)}{\expop[w(X_l)]}
\label{eq:entropyRate}
\end{align}
where $\expop[w(X_l)]$ denotes the average weight of all $x\in\mathcal{X}_l$ with
respect to the PMF $p_{X_l}$ and where $\entop(X_l)$ denotes the
entropy of $X_l$ in nats.
\end{definition}
We can upper-bound the entropy rate $\bar{\entop}(X)$ by maximizing each term of the sequence on the
right-hand side of \eqref{eq:entropyRate} separately. To do so, we need the following lemma:
\begin{lemma}
\label{lem:entropyBound}
Denote by $p_{Z}$ the PMF of some random variable $Z$ with countable support $\mathcal{Z}$ and an associated positive weight function $w$. The maximum entropy per average weight
\begin{align}
 R_{Z}&=\max_{p_{Z}}\frac{\entop(Z)}{\expop[w(Z)]}
\label{eq:maximumEntropyRate}
\end{align}
is given by the greatest positive real solution of the equation
\begin{align}
\sum\limits_{z\in\mathcal{Z}}e^{-w(z)s}=1.
\label{eq:maximumEntropie}
\end{align}
In addition, the PMF of $Z$ that achieves this
rate is uniquely given by
\begin{align}
q_{Z}(z)=e^{-w(z)R_{Z}},\quad z\in\mathcal{Z}.
\end{align}
\end{lemma}
\begin{IEEEproof}
These two properties of $R_{Z}$ were derived by using Lagrange Multipliers
in \cite{Marcus1957} and they were independently derived in \cite{Krause1962} by
using the bound
$\ln z \leq z-1$. We offer an alternative proof by applying the information
inequality \cite{Cover2006}, which states for the Kullback Leibler Distance
$D(\cdot\Vert\cdot)$ of two PMFs $p$ and $q$ that
\begin{align}
 D(p\Vert q)\geq 0
\end{align}
with equality if and only if $p=q$. We thus have
\begin{align}
 0&\geq -D(p_{Z}\Vert q_{Z})\\
&=\sum\limits_{z\in\mathcal{Z}}p_{Z}(z)
\ln\frac{q_{Z}(z)}{p_{Z}(z)}\\
&=\entop(Z)-R_{Z} \expop[w(Z)]
\end{align}
which implies
\begin{align}
\frac{\entop(Z)}{\expop[w(Z)]}\leq R_{Z}
\end{align}
with equality if and only if $p_{Z}=q_{Z}$.
\end{IEEEproof}
\begin{lemma}
\label{lem:rateBound}
Let $X$ denote an input source of some constrained system. Let the \emph{rate bound} $\mathsf{R}_X$ be defined as
\begin{align}
\mathsf{R}_X=\limsup_{l\rightarrow\infty}R_{X_l}\label{eq:def:rateBound}
\end{align}
where each $R_{X_l}$ is chosen according to Lemma~\ref{lem:entropyBound}. The entropy rate $\bar{\entop}(X)$ of $X$ is is then upper-bounded by $\mathsf{R}_X$.
\end{lemma}
\begin{IEEEproof}
We have
\begin{align}
\bar{\entop}(X)&=\limsup_{l\rightarrow\infty}\frac{\entop(X)}{\expop[w(X_l)]}\\
&\leq\limsup_{l\rightarrow\infty}\max_{p_{X_l}}\frac{\entop(X)}{\expop[w(X_l)]}\\
&=\limsup_{l\rightarrow\infty}R_{X_l}\\
&=\mathsf{R}_X.
\end{align}
\end{IEEEproof}
\begin{lemma}
\label{lem:bound}
Let $A=(\mathcal{A},w)$ represent a constrained system and let $X$ denote an input source of $A$. The rate bound $\mathsf{R}_X$ of $X$ is then upper-bounded by the abscissa of convergence $Q$ of $\gf_A$, i.e.,
\begin{align}
\mathsf{R}_X\leq Q.
\end{align}
\end{lemma}
\begin{IEEEproof}
To proof the lemma, we show that the generating function $\gf_A(s)$ diverges whenever
$\repart(s)<\mathsf{R}_X$, for any input source $X$.

The definition of the rate bound $\mathsf{R}_X$ in \eqref{eq:def:rateBound} implies in particular that for any $\epsilon>0$
\begin{align}
R_{X_l}&\geq \mathsf{R}_X-\epsilon\quad\text{i.o.}
\end{align}
According to \eqref{eq:maximumEntropie}, $R_{X_l}$ is given by the greatest positive real solution of
\begin{align}
\sum\limits_{x\in\mathcal{X}_l}e^{-w(x)s}=1
\end{align}
which implies further
\begin{align}
 \sum\limits_{x\in\mathcal{X}_l}e^{-w(x)[\mathsf{R}_X-\epsilon]}&\geq
\sum\limits_{x\in\mathcal{X}_l}e^{-w(x)R_{X_l}}=1\quad\mathrm{i.o.}
\label{eq:probabilitySumBoundIof}
\end{align}
Because of $\bigcup_{l=1}^\infty \mathcal{X}_l\subseteq\mathcal{A}$ according to Definition~\ref{def:inputSource}, we can bound the generating function by
\begin{align}
 \gf_A(s)=\sum\limits_{a\in \mathcal{A}}e^{-w(a)s}\geq\lim_{n\rightarrow\infty}
\sum\limits_{l=1}
^n\sum\limits_{x\in\mathcal{X}_l} e^ {-w(x)s}.
\end{align}
Because of \eqref{eq:probabilitySumBoundIof}, we have
\begin{align}
 \sum\limits_{l=1}
^n\sum\limits_{x\in\mathcal{X}_l}e^{-w(x)[\mathsf{R}_X-\epsilon]}\overset{n\rightarrow\infty}{
\longrightarrow}\infty
\end{align}
and we conclude that for any $\epsilon>0$, $\gf_A$ diverges in $s=\mathsf{R}_X-\epsilon$. Thus, by \cite[Theorem 3]{Hardy1915}, $\gf_A$ diverges for all $s\in\mathbb{C}$ with $\repart(s)<\mathsf{R}_X$. Since by definition of $Q$, $\gf_A$ converges for $\repart(s)>Q$, it must hold that $\mathsf{R}_X\leq Q$.
\end{IEEEproof}

%% file: maxentropic.tex
\section{Maxentropic Input Processes}
\label{sec:inputProcess}
We now come to the main concern of this work: we want to define input processes for constrained systems and we want to investigate how the entropy rate of an input process is related to the combinatorial capacity. Loosely speaking, we want to define a random process that generates at each time instant a substring, which is then appended to the string that has been generated so far. At each time instant, the complete string should fulfill the constraints of the considered system. The notion of an input process differs fundamentally from what we defined as an input \emph{source}: an input source generates a complete new string at each time instant. Before we give our definition of input processes, we motivate our definition by the following example. Consider a constrained system $S_\mathrm{bin}$ that accepts any binary sequence, and assume $w(\strone)=w(\strzero)=1$. The combinatorial capacity is $\mathsf{C}=\ln(2)\approx 0.6932$. Denote by $V=\{V_l\}_{l=1}^\infty$ the random process where the $V_l$ take values in $\{\strzero,\strone,\strzero\strone\}$ and are independent, identically distributed (IID) according to the PMF $p_V$, which we define as follows:
\begin{align}
p_V(\strzero)=p_V(\strone)=e^{-R},\quad p_V(\strzero\strone)=e^{-2R}
\end{align}
where $R$ is given by the largest positive real solution of
\begin{align}
2e^{-s}+e^{-2s}=1.
\end{align}
Obviously, $V$ generates binary strings that are accepted by $S_\mathrm{bin}$.
We are interested in the entropy rate of $V$ and calculate
\begin{align}
\frac{\entop(V)}{\expop[w(V)]}&=R\label{eq:entropyAttempt}\\
&\approx 0.8814\\
&>\mathsf{C}
\end{align}
where equality in \eqref{eq:entropyAttempt} follows from Lemma~\ref{lem:entropyBound}. Surprisingly, the entropy rate of $V$ seems to exceed the combinatorial capacity of $S_\mathrm{bin}$. The reason for this is that we implicitly assume in our attempt to calculate the entropy rate of $V$ that, for example, the realizations $v_1=\strzero\strone$ and $(v_1,v_2)=(\strzero,\strone)$ are distinguishable, however, they are not: both result in the string $\strzero\strone$, so we are counting this string twice. To avoid this pitfall, we define input processes as follows.
\begin{definition}
The random process $\{Y_l\}_{l=1}^\infty$, $Y_l\in\mathcal{Y}$ is an \emph{input process} of the constrained system $A=(\mathcal{A},w)$ if the sequence of random variables $X_l=\catop(Y_1,\dotsc,Y_l)$ with the supports truncated to
\begin{align}
\mathcal{X}_l=\left\{\catop(y_1,\dotsc,y_l)\vert (y_1,\dotsc,y_l)\in\mathcal{Y}^l,\;p_Y(y_1,\dotsc,y_l)>0
\right\}\label{eq:truncatedSupport}
\end{align}
is an input source of $A$. The operator $\catop$ denotes concatenation: \text{$\catop(\stra,\strb)=\stra\strb$}.
\end{definition}
We refer to the assignment \eqref{eq:truncatedSupport} in the following by \emph{truncated support}.

The process $V$ as we defined it earlier is not an input process of $S_\mathrm{bin}$: define $X_l=\catop(V_1,\dotsc,V_l)$, $l=1,2,\dotsc$. The random variables $X_1$ and $X_2$ have the following truncated supports: 
\begin{align}
\mathcal{X}_1&=\{\strzero,\strone,\strzero\strone\}\\
\mathcal{X}_2&=\{\strzero\strzero,\strzero\strone,\strzero\strzero\strone,
\strone\strzero,\strone\strone,\strone\strzero\strone,
\strzero\strone\strzero,\strzero\strone\strone,\strzero\strone\strzero\strone\}.
\end{align}
As we can see, $\mathcal{X}_1\cap\mathcal{X}_2=\strzero\strone\neq\emptyset$, so $X$ is not an input source of $S_\mathrm{bin}$ and thus, by definition, $V$ is not an input process of $S_\mathrm{bin}$. By changing the PMF of the $V_l$ to
\begin{align}
p_V(\strzero)=p_V(\strone)=\frac{1}{2},\quad p_V(\strzero\strone)=0\label{eq:inputProcessProbabilities}
\end{align}
each random variable $X_l$ has the truncated support
\begin{align}
\mathcal{X}_l = (\strzero\cup\strone)^l.
\end{align}
Thus, with the probability assignment \eqref{eq:inputProcessProbabilities}, $\mathcal{X}_l\cap\mathcal{X}_k=\emptyset$ whenever $k\neq l$, and consequently, $X$ is an input source and $Y$ is an input process.

Now that we have defined input processes in a way that does not allow for ``counting twice'', we can give the following
\begin{definition}
The \emph{entropy rate} of an input process $Y$ is defined as
\begin{align}
\bar{\entop}(Y)=\limsup_{l\rightarrow\infty}\frac{\entop(Y_1,\dotsc,Y_l)}{\expop[w(Y_1)+\dotsb+w(Y_l)]}.\label{eq:entropyRateProcess}
\end{align}
\end{definition}
The entropy rate of an input process of a constrained system relates to the combinatorial capacity as follows.
\begin{theorem}
\label{theo:inputProcess}
Let $A=(\mathcal{A},w)$ represent a constrained system. The entropy rate of an input process $Y$ of $A$ is upper-bounded by the abscissa of convergence $Q$ of $\gf_A$, and in particular, it is upper bounded by the combinatorial capacity $\mathsf{C}$ of $A$.
\end{theorem}
\begin{IEEEproof}
Since $Y$ is an input process of $A$, by definition, the sequence $X$ of random variables $X_l=(Y_1,\dotsc,Y_l)$ with truncated supports is an input source of $A$. We thus have
\begin{align}
\bar{\entop}(Y)&=\limsup_{l\rightarrow\infty}\frac{\entop(Y_1,\dotsc,Y_l)}{\expop[w(\catop(Y_1,\dotsc,Y_l))]}\\
&= \limsup_{l\rightarrow\infty}\frac{\entop(X_l)}{\expop[w(X_l)]}\label{eq:propProcess:limsup}\\
&\leq\mathsf{R}_X\label{eq:propProcess:rateBound}\\
&\leq\mathsf{Q}\label{eq:propProcess:convergence}\\
&=\mathsf{C}\label{eq:propProcess:combinatorial}
\end{align}
where the equality in \eqref{eq:propProcess:limsup} follows from the definition of $X$, the inequality in \eqref{eq:propProcess:rateBound} follows from Lemma~\ref{lem:rateBound}, the inequality in \eqref{eq:propProcess:convergence} follows from Lemma~\ref{lem:bound}, and the equality in \eqref{eq:propProcess:combinatorial} follows from Theorem~\ref{theo:capacityConvergence}.
\end{IEEEproof}
%
%
\begin{figure}
\setcounter{textequation}{\value{equation}}
\renewcommand{\theequation}{E.\arabic{equation}}
\setcounter{equation}{\value{exampleequation}}
\begin{framed}
Let $Y=\{Y_l\}_{l=1}^\infty$ denote a random process with realizations $(y_1,\dotsc,y_l)\in\mathcal{Y}^l$. Let $\mathcal{Y}$ be given by
\begin{align}
\mathcal{Y}=(\strzero\cup\dotsb\cup \underbrace{\strzero\dotsb \strzero}_\text{$k$ times})(\strone\cup\dotsb\cup \underbrace{\strone\dotsb \strone}_\text{$j$ times}).
\end{align}
From Example~\ref{fig:inputSource}, we know that the sequence of random variables $\{X_l\}_{l=1}^\infty$ with $X_l=\catop(Y_1,\dotsc,Y_l)$ is an input source of $S_{(j,k)}$; therefore, $Y$ is an input process of $S_{(j,k)}$. 

Let $\{Y_l\}_{l=1}^\infty$ be independent, identically distributed (IID) according to the PMF
\begin{align}
p_{Y}(y)=e^{-w(y)R},\quad y\in\mathcal{Y}
\end{align}
where $R$ is given by the largest positive real solution of
\begin{align}
\sum\limits_{y\in\mathcal{Y}}e^{-w(y)s}&=(e^{-s}+\dotsb+e^{-js})(e^{-s}+\dotsb+e^{-ks})\nonumber\\
&=1.\label{eq:exampleEntropyRate}
\end{align}
Comparing \eqref{eq:exampleEntropyRate} with \eqref{eq:exampleCapacity} in Example~\ref{fig:combinatorialCapacity}, we see that $R=\mathsf{C}$, i.e., $R$ is equal to the combinatorial capacity $\mathsf{C}$ of $S_{(j,k)}$. 
The entropy rate of $Y$ is
\begin{align}
\bar{\entop}(Y)&=\limsup_{l\rightarrow\infty}\frac{\entop(Y_1,\dotsc,Y_l)}{\expop[w(\catop(Y_1,\dotsc,Y_l))]}\\
&=\limsup_{l\rightarrow\infty}\frac{l\entop(Y)}{l\expop[w(Y)]}\\
&=R=\mathsf{C}
\end{align}
where the second equality follows from the independence bound on entropy \cite{Cover2006} together with the fact that the $Y_l$ are IID, and the linearity of $w$. Since, from Theorem~\ref{theo:inputProcess}, $\bar{\entop}(Y)\leq\mathsf{C}$, we conclude that $Y$ is a maxentropic input process of $S_{(j,k)}$.
\end{framed}
\vspace{-0.4cm}
\caption{A maxentropic input process of $S_{(j,k)}$}
\vspace{-0.4cm}
\label{fig:inputProcess}
\setcounter{exampleequation}{\value{equation}}
\setcounter{equation}{\value{textequation}}
\end{figure}
%
%
We call an input process $Y^*$ of a constrained \text{system $A$} \emph{maxentropic} if for any input process $Y$ of $A$ we have $\bar{\entop}(Y^*)\geq\bar{\entop}(Y)$. Because of Theorem~\ref{theo:inputProcess}, $\bar{\entop}(Y^*)=\mathsf{C}$, i.e., the entropy rate of $Y^*$ is equal to the combinatorial capacity of $A$, is a sufficient condition for $Y^*$ to be maxentropic. It is important to note that Theorem~\ref{theo:inputProcess} does not claim that $\bar{\entop}(Y^*)=\mathsf{C}$ for any constrained system $A$. For a large class of constrained systems, however, $\bar{\entop}(Y^*)=\mathsf{C}$. For this class, Theorem~\ref{theo:inputProcess} is quite useful: assume that we want to show that system $A$ belongs to this class. Without Theorem~\ref{theo:inputProcess}, we have to maximize the entropy rate over all input processes of $A$. Once we have determined the maximum entropy rate, we compare it to the combinatorial capacity and find that both are equal. An example of this approach can be found in the proof of \cite[Theorem 5.1]{Khandekar2000}. With Theorem~\ref{theo:inputProcess}, we can do something different: we look for an input process whose entropy rate is equal to the combinatorial capacity. Once we have found such an input process, we invoke Theorem~\ref{theo:inputProcess} and are done. We illustrate this new approach in Example~\ref{fig:inputProcess}.

%% file: conclusions.tex
\section{Conclusions}
In this work, we showed for a general class of constrained systems (including those with non-regular constraints and dropping the ``not too dense'' assumption for the weight set) that the maximum entropy rate of input processes is upper-bounded by the combinatorial capacity of the considered system. This general result allows for a new approach to show that maximum entropy rate and combinatorial capacity are \emph{equal}: with our result, it is enough to find ``some'' input process whose entropy rate is equal to the combinatorial capacity of the considered system. Equality of \emph{maximum} entropy rate and combinatorial capacity then follows from our result. In contrast to previous works (except for some works that consider specific classes of constraint systems), we do not use any result from matrix theory in our derivations. Our framework, which is based on generating functions, therefore allows, besides being more general, for a new approach to investigate regular systems. We illustrated this by applying our results to the $(j,k)$ constraint.